\documentclass[journal]{vgtc}                     

\onlineid{0}

\usepackage{tabularx}
\usepackage{balance}

\vgtccategory{Research}

\title{Assessing Affective Objectives for Communicative Visualizations}

\author{%
  \authororcid{Elsie Lee-Robbins}{0000-0002-4080-6506}, and
\authororcid{Eytan Adar}{0000-0003-1911-836X}
}

\authorfooter{
  \item
  	Elsie Lee-Robbins was with the University of Michigan while completing this work, Eytan Adar is at the University of Michigan
}

\abstract{%
Using learning objectives to define designer intents for communicative visualizations can be a powerful design tool. Cognitive and affective objectives are concrete and specific, which can be translated to assessments when creating, evaluating, or comparing visualization ideas. However, while there are many well-validated assessments for cognitive objectives, affective objectives are uniquely challenging. It is easy to see if a visualization helps someone remember the number of patients in a clinic, but harder to observe the change in their attitudes around donations to a crisis. In this work, we define a set of criteria for selecting assessments--from education, advocacy, economics, health, and psychology--that align with affective objectives. We illustrate the use of the framework in a complex affective design task that combines personal narratives and visualizations. Our chosen assessments allow us to evaluate different designs in the context of our objectives and competing psychological theories.
}


\teaser{
  \centering
  \includegraphics[width=\linewidth, alt={Examples of several affective visualizations. (A) Freeing the US from its culture of detention as a waffle chart, from Voila. (B) an arc diagram of US Gun Deaths, from Periscopic \cite{Periscopic_2013}. (C) Climate change knows no borders, from Gabrielle Merite with a data visualization and small images of people. (D) Two screenshots of Abandoned at Sea, from Kontinentalist in partnership with UN HCR. The top screenshot shows an illustration of a refugee walking towards a boat, part of a narrative, and an audio file from Asmotulleh, a refugee. The bottom shows a map of refugees abandoned at sea.}]{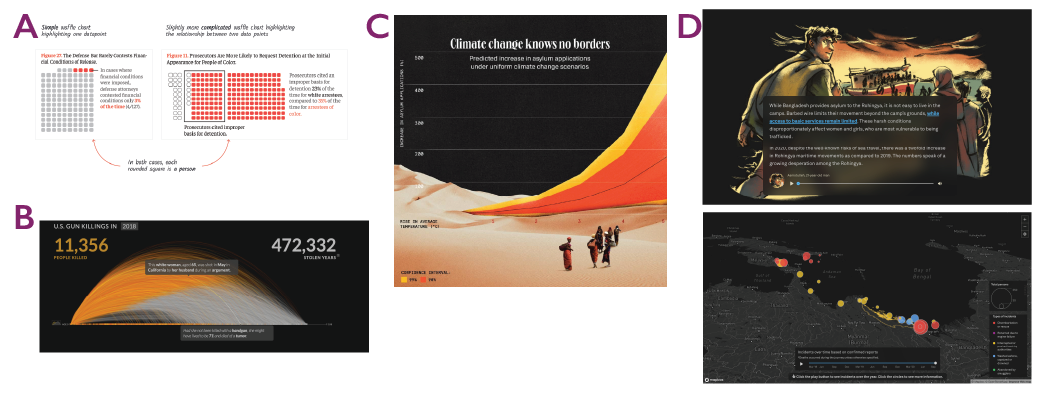}
  \caption{%
  	Examples of several affective visualizations. (A) Freeing the US from its culture of detention: waffle chart, from Voila \cite{Voila2023}. (B) US Gun Deaths, from Periscopic \cite{Periscopic_2013}. (C) Climate change knows no borders, from Gabrielle Merite \cite{Merite_2021}. (D) Two screenshots of Abandoned at Sea, from Kontinentalist in partnership with UN HCR \cite{Kontinentalist_2020}. The top screenshot shows an illustration of a refugee walking towards a boat, part of a narrative, and an audio file from Asmotulleh, a refugee. The bottom screenshot shows a map of refugees abandoned at sea. 
  }
  \label{fig:teaser}
}

\graphicspath{{figs/}{figures/}{pictures/}{images/}{./}} 

\usepackage{tabu}                      
\usepackage{booktabs}                  
\usepackage{lipsum}                    
\usepackage{mwe}                       
\usepackage{ccicons}                   

\usepackage{mathptmx}                  

\begin{document}

\firstsection{Introduction}

\maketitle

The use of learning objectives (LO) in the construction of communicative visualization~\cite{adar2021,lee2022affective} can help designers better create and select visualizations~\cite{leerobbins22}. Well-designed LOs have the benefit of being translatable to assessments~\cite{guan25}. Thus, a visualization designer can clearly state their intent--e.g., \textit{the viewer will recall the unemployment levels in the US}--and create an appropriate assessment. For example, a simple multiple choice question: \textit{What is the current unemployment level in the US? A) 5.1\%, B) 4.3\% or C) 4.6\%}. By testing a viewer pre-- and post-exposure to one or more visualizations, the designer can be more confident that their intents were achieved. 

Learning objectives come in two primary flavors: cognitive and affective, and both can be used to define visualization intents (a third category, psychomotor, is harder to align). Cognitive objectives are easier to describe~\cite{anderson2001taxonomy} using the standard \textbf{the viewer will [verb] [noun]} form. The verbs--recall, critique, summarize, etc.--are directly mappable to the cognitive changes the designer intends. Similarly, the nouns (specific facts, concepts, etc.) are more easily connected to data. The clearer definitions of these constructs are a key part of why we can more easily construct \textit{cognitive} assessments.

However, the intent of many communicative visualizations goes beyond the cognitive to the affective. For example, advocacy groups and organizations with a mission to improve the world often use data collected about humanitarian crises (see Figure~\ref{fig:teaser}). Their goals might be to raise awareness of the crisis and elicit responses from their audience: donate money to the cause, share the message with friends and family, or contact policymakers. Expressed as affective LOs: they aim to influence the audience's appraisals, attitudes, or values~\cite{lee2022affective}. 

In many ways, designing communicative visualizations for \textit{cognitive} goals is more straightforward. There is significant guidance on their construction, significant research on their effectiveness, and there are rarely alternatives to visualizations. Cognitive goals are so entangled with the data itself that it is hard to argue that visualizations are not nearly always optimal for this task. With affective goals, there is far less guidance and research. Designing them well is important as the use of visualizations and data can actually backfire when design goals are affective. Impersonal charts or a focus on large population statistics, for example, can lead to numbing~\cite{Slovic_Slovic_2015}. However, the answer to simply abandon visualizations is rarely the right one. Visualizations are often a response to a wicked design problem where multiple objectives--mixtures of cognitive and affective--must be balanced. In some situations designers can augment their designs with non-visualization features (either exclusively or in combination with visualizations). The key to making these decisions is to have good assessment tools. Intuition, even combined with specific design guidelines driven by well--studied theories, is unlikely to tell the designer which visualization design will work best or better. 

The issue we tackle in this work is that the guidance on properly assessing \textit{affective} LOs is very limited. Within education, affective assessments often focus on the class (e.g., did you like the class? do you want to take another math class? etc.) and suffer from various validity issues. Fortunately, experimental evaluation/measures from other domains (e.g., economics, advocacy, marketing, psychology, etc.) do offer alternatives.
By restricting ourselves to assessments within the design process, we emphasize the need for timely (i.e., scalable and easy to run), actionable (that allow for contrasting design alternatives), and well-validated assessments. Our goal in this work is to articulate these constraints as guidelines for selecting assessments.  Through the guidelines, we can dismiss assessment approaches where constructs or other implementation challenges make them unsuitable for our context. For example, certain political constructs are too vague, psychological scales may be unreliable, and advocacy evaluations can require time frames that greatly exceed a visualization design cycle. On the other hand, we can reclaim instruments which may be unreliable in certain environments (e.g., self-reports in a classroom settings) because they are reasonable in our setting. 

\begin{figure}[t]
     \centering
  \includegraphics[width=\linewidth, alt={}]{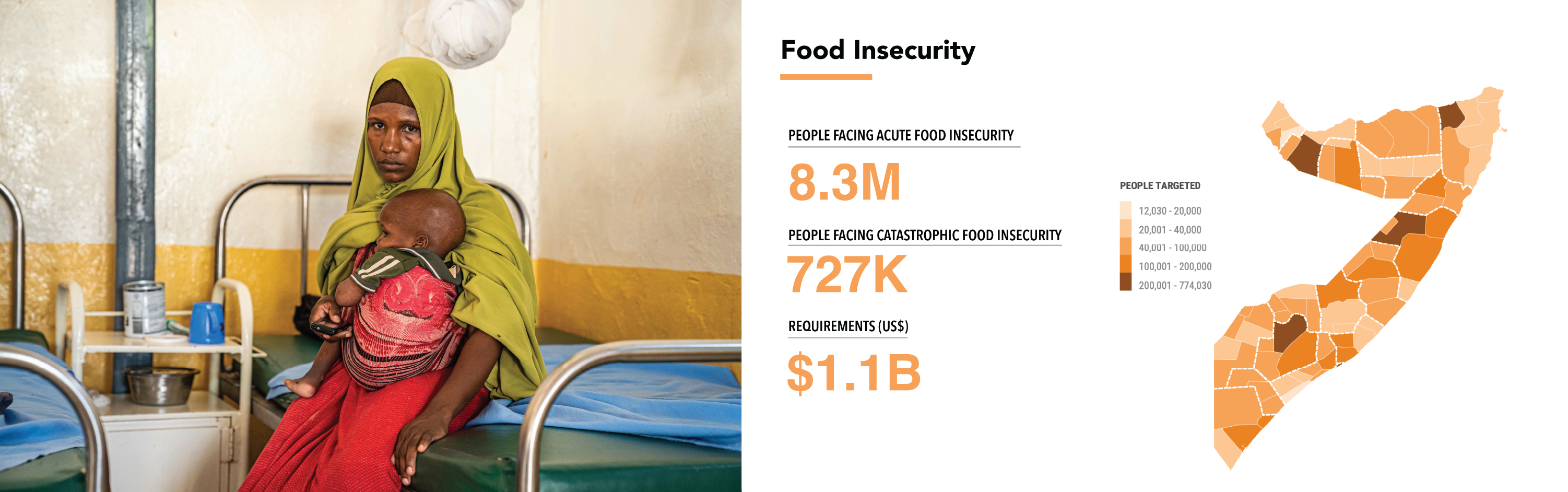}
  \caption{%
  	(Left) A picture of Ruqio, a Somali woman, and her child in a hospital. Image from the United Nations Office for the Coordination of Humanitarian Affairs (UN OCHA) \cite{UNOCHA_2022}. (Right) Data visualization of the population and location of Somalis facing food insecurity. Chart recreated, based on UN OCHA's Humanitarian Response Plan Report \cite{UNOCHA_2023}. 
  } 
  \label{fig:advocacy}
\end{figure}

We demonstrate how to utilize different affective assessments within a realistic case study. In the example, the designer has a mix of affective objectives (e.g., \textit{believe in the importance of the humanitarian efforts}, \textit{act to donate}, etc.). To achieve these, the designer can apply a mix of rhetorical and design strategies. They may leverage a logical argument (e.g., the rhetorical \textit{logos}) and use statistics, numbers, and charts to draw a compelling picture of an issue by presenting data as evidence. Numbers and statistics can be presented as facts, wielded as reasons to support a persuasive argument. Data visualization is a tool that can effectively show trends, patterns, and the scope of problems (Figure~\ref{fig:advocacy}, right). Alternatively, they may emphasize an emotional appeal (the rhetorical \textit{pathos}) by showing photographs, audio, videos, and stories about individuals. For example, photography is often the most popular choice for certain types of advocacy communication~\cite{rall2016data} but other types of illustration and visual art are also common (Figure~\ref{fig:advocacy}, left). Using the LO formulation and assessment approaches, we demonstrate how different communicative alternatives can be compared. 

This work contributes guidance on how affective objectives can be measured. Specifically, we describe the unique constraints of the assessment of affective visualization and a set of evaluation suggestions for a wide range of affective objectives. We offer a specific example of how proper assessment can inform a designer challenged by multiple objectives and conflicting research evidence.

\section{Background}
\label{section:related}

\subsection{Affective Visualizations}

Affective visualizations have gained more attention as researchers consider non-cognitive intents~\cite{lan2023affective,lee2022affective}. The term \textit{affective} has several definitions in this community. In this work, we utilize the definition of Lee-Robbins and Adar~\cite{lee2022affective}) that treats affective goals as those that aim to influence appraisals, attitudes, values, or value systems (broadly, \textit{beliefs}). There are many types of beliefs that can be considered under this definition, and a clear articulation is challenging. However, for our purposes, we consider those constructs where (a) they can be learned or changed, and (b) where different people might hold different stances (e.g., positive or negative) on those beliefs. Defining these by exclusion, we \textit{do not consider} affective targets that are primitive (non-reliant on other beliefs) and for which there is 100\% consensus (everyone agrees)~\cite{Rokeach68} (e.g., beliefs that are ``psychologically incontrovertible''). Succinctly, these are ideas where not everyone agrees on the same thing and there is an opportunity to change a person's mind.

Affective visualizations (see Figure~\ref{fig:teaser}) are visualizations created to achieve one of these goals. A specific category for affective visualizations is anthropographics, which have been defined as ``\textit{visualizations that represent data about people in a way intended to promote prosocial feelings (e.g. compassion or empathy) or prosocial behavior (e.g., donating or helping)}''~\cite{Morais_Jansen_Andrade_Dragicevic_2020}. Affective communicative visualizations can exist for many reasons, but they are often applied as a form of advocacy--pushing the viewer towards having some negative (or positive) view of some belief or assertion and behaving accordingly. Because advocacy often involves various rhetorical strategies, this area offers many examples of evaluation approaches of affective visualizations. 

Advocacy-focused goals are a useful representation of affective communicative goals (and consequently affective visualization goals). At their core, advocacy aims to help other people. For example, an advocacy organization can launch advertising campaigns to raise awareness about an issue. Often, they organize fundraising drives to get donations for their cause. Non-visualization approaches may be to present testimonials, stories, and in-depth information about individuals. Individual photographs (e.g., of a person impacted) are often effective tools. People's stories and photographs capture attention and can evoke an emotional connection (a \textit{rhetorical pathos} strategy)~\cite{bogre2012photography}. 

An alternative approach, which we focus on, is to make a more data-driven appeal to the audience~\cite{Tactile_Technology_Collective_2013, ganesh2015communicating}. Nonprofit and government organizations have increasingly used data visualizations in advocacy reports in recent decades~\cite{rall2016data}. There are several reasons why advocacy groups might want to present data visualizations. First, data could give advocacy groups credibility and trustworthiness~\cite{rall2016data} (an \textit{ethos} rhetorical strategy). Data provides a logical argument and `evidence' to support claims. Second, some designers perceive data visualizations to be persuasive~\cite{Pandey_Manivannan_Nov_Satterthwaite_Bertini_2014, rall2016data} (a \textit{logos} strategy). Third, data visualizations could simplify and distill complex information and make it accessible to a wider audience~\cite{ganesh2015communicating}.

Outside of advocacy, we see examples of affective visualizations in marketing, political propaganda, or health communication. In all these scenarios visualizations act as rhetorical objects, often targeting both cognitive and affective channels. Their objective is to convince the viewer of something and have them act accordingly--to buy, to vote, to take care of their health. Arguably, most visualizations, even those that aspire for neutrality or unbiased presentation (e.g., a visualization in a newspaper article or academic presentation) have some implicit affective objectives, often attempting to convince the viewer of the `correctness' of an argument~\cite{lee2022affective}. Assessing whether these affective objectives have been met is a key focus of this work.

\subsection{Affective Learning Objectives}
In this work, we leverage our previously-defined LO taxonomy~\cite{lee2022affective}. The taxonomy offers a classification of verbs and nouns in the standard framing: \textit{the viewer will [verb] [noun]}. This differs in a number of ways from the traditional affective objective taxonomy of Krathwol et al.~\cite{krathwohl1956taxonomy}, which lacked concrete structure for the `nouns' and did consider a broader set of objectives beyond those in a classroom setting. It is also worth noting that we take a broad understanding of `affective' to include not just emotions and feelings. In the case of cognitive objectives, \textit{perception} (e.g., reading the height of a bar in a bar chart) is necessary but not sufficient for many objectives (e.g., recalling something's value). Similarly, for affective objectives, an emotional response is often insufficient (and may, or may not, be necessary). Rather, a designer will seek to encourage changes in beliefs, attitudes, or values that persist beyond the initial emotional reaction. More critically, an emotion-driven strategy (i.e., pathos) may not also be appropriate to achieve an affective rhetorical objective.

\begin{figure*}[t]
     \centering
  \includegraphics[width=.95\linewidth, alt={}]{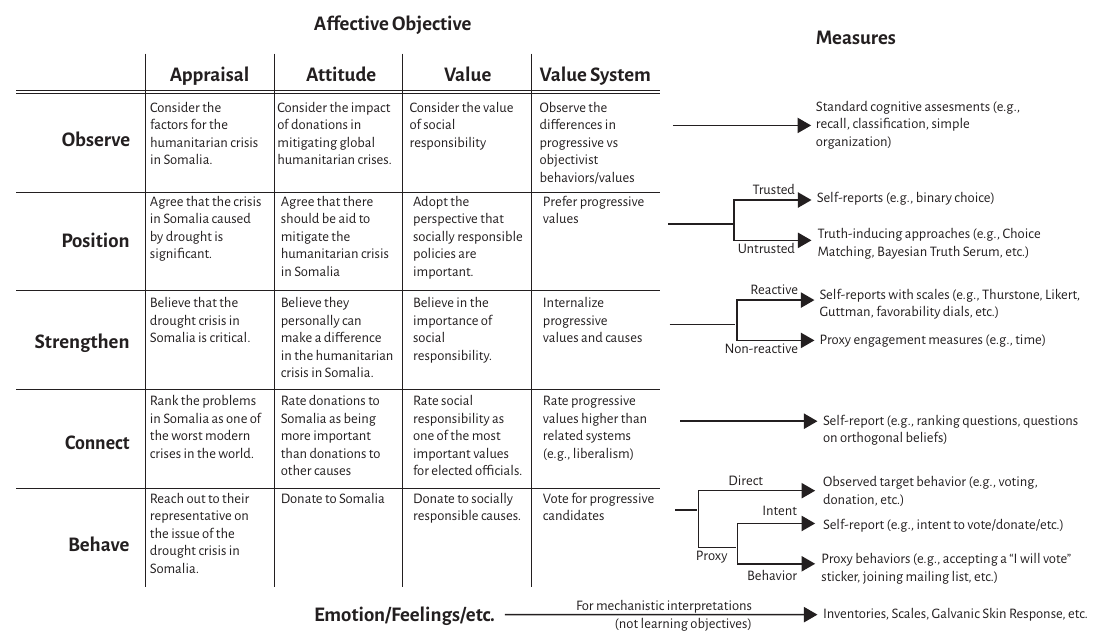}
  \caption{Structure for affective learning objectives in communicative visualizations with possible measures. Each cell in the table (left) provide an example objective.}
  \label{fig:table}
  
\end{figure*}

Figure~\ref{fig:table} (left) summarizes the noun/verb grid for the taxonomy with examples. We note a few specific attributes of the taxonomy in that they impact our assessment choices. The verbs (rows) focus on what we want the viewer to be able to do after being exposed to the visualization. 

This affective objective structure defines the space for which we would like to identify assessments. As with most `learning' situations, a designer might have multiple objectives that are a mix of affective and cognitive and may be in opposition to each other as part of a wicked design problem~\cite{buchanan1992wicked}. This will necessarily impact design choices and implicate assessment strategies. Additionally, as with cognitive objective frameworks~\cite{anderson2001taxonomy}, verbs and nouns are only weakly hierarchical~\cite{haladyna2004developing}. For example, some verbs (e.g., \textit{Position}) may depend on others (e.g., \textit{Observe}). However, it is not necessary to achieve a \textit{Connect} objective before targeting a \textit{Behave} one. Again, this will have an impact on how assessments are structured, as it is possible that achieving a `high-level' objective such as \textit{Behave} or \textit{Connect} does not mean the \textit{Strengthen} objective has been met. 

\subsection{Affective Assessment in Education}
In theory, the advantage of creating a well-defined set of learning objectives should mean that these are translatable to assessments that can cover a wide range of topics. In the case of cognitive learning objectives, this has certainly happened as instructors have created questions, tests, and rubrics for all types and topics. Unfortunately, the same cannot be said for affective objectives. 

Taxonomies for affective objectives have evolved over time, but nowhere near the rate of other objective types~\cite{martin1986affective,krathwohl1956taxonomy}. Unlike Bloom's original cognitive taxonomy, which has seen continuous refinement and updating~\cite{anderson2001taxonomy}, affective objectives have remained more fixed. In part, this is due to the difficulty of assessment. The 1956 form of the taxonomy which includes a hierarchy of verbs (\textit{receive}, \textit{respond}, \textit{value}, \textit{organize}, and \textit{characterizing}) but no principled model of the nouns to which they should be applied. The bulk of examples from affective education target a specific set of objectives that are often about the act of learning itself (i.e., `the student will appreciate math', `the student will register for more English classes') or classroom behaviors (`the student will not bully others')~\cite{anderson2013assessing,cook2016motivation}.

A complete discussion of why education has shied away from broader engagement with affective objectives is beyond the scope of this paper. However, we consider a few key aspects of this legacy. First, there are often ethical concerns for affective objectives. That these may reflect advocacy or even brainwashing~\cite{martin1986affective} and a deviation from the `knowledge credentialing' focus was inappropriate. Outside the classroom, as communicative visualizations often are, this concern is more limited~\footnote{Though this does not mean that affective objectives are uncontroversial in the information visualization community (see~\cite{lee2022affective} for a broader discussion).}.

A second issue with affective objectives in educational contexts is that they are often assessed anonymously and/or without grading~\cite{martin1989checklist} and thus are unreliable. Although one could reasonably construct a test to assess math knowledge or a rubric to evaluate a written document, affective objectives are much more challenging. In the case of self-reports, which is the predominant educational assessment mechanism, the test taker often recognizes the social-- and teacher-desirable answer and will respond accordingly. In the context of communicative visualizations outside the class, these issues may be less important. The viewers are not incentivized to produce socially-desirable responses and more incentive-compatible and elicitation protocols can be used.

Finally, classroom assessments are often primarily aimed at evaluating individual students rather than determining programmatic value (i.e., how much the teacher, textbook, activity, or module helped). That is, the goal is to determine an individual student's performance either relative to others in the class or to some baseline. In our context, assessments are entirely focused on the program and can easily produce and evaluate alternatives (e.g., `does the visualization help achieve the objective?' and `which visualization is better?'). It is much easier to produce alternative designs and find an audience for evaluation. This will make many other types of assessments available to us.

\section{Selecting and Constructing Assessments}
\label{section:assessments}
In identifying assessments for affective visualizations, it is critical to ask: what is unique in the \textit{design process} of communicative visualizations? Primarily, we are interested in those time-bound activities in which a designer produces an artifact that contains visualizations. These visualizations may be static elements for an article, an infographic, a video, a presentation, or interactive treatments for the web. In most projects, the designer will have some mix of cognitive and affective objectives--some of which may be in conflict with others or be prioritized differently. Visualization design under these constraints is often a \textit{wicked design} problem~\cite{buchanan1992wicked}. A solution may never be perfect or optimal, but may be produced as part of some tradeoffs. `Mastery' of any particular objective may never be achieved. In most cases, the designer is seeking to do the best they can towards satisfying their objectives under constraints. Assessments offer a mechanism by which they can determine progress.

\subsection{What's better?}
In its simplest form, an assessment simply resolves to some operationalized measure of learning, or in our case, belief. In many situations, this treats the visualization as an intervention where the same question can be asked before and after exposure (during or after showing the visualization): How important is this topic before I showed you the data? and how important do you believe it is after?  Naturally, the designer might have many different objectives, each with their own indicator. Assuming that no visualization `dominates' all objectives, this would mean making a tradeoff of which objectives are most important.

Not all assessments require both pre-- and post-tests. When comparing alternatives, the designer may be satisfied with showing different people (e.g., half the participant population) the different options and comparing their responses\footnote{A more statistically sophisticated version of this would be the difference-in-difference (DID)~\cite{abadie2005semiparametric} which could control for differences in the population looking at visualization 1 and those in visualization 2}. The advantage of visualization design is that design variations are often a natural output of the design process. During the design process, the designer can produce alternative designs that they intuitively believe will better serve their learning objectives: transforming the data, changing encodings, adding visual embellishments, creating specific annotations, modifying colors, etc. Thus, the act of design will invariably produce alternative forms.

We note that a between-subjects strategy is likely preferable for the assessments we describe here. That is, each participant only sees one version of the design. Clearly, exposing a viewer to multiple visualizations in a row might create affective priming from the first visualization they see, impacting the results for the second.

As we describe below, each domain (education, marketing, politics, non-profit advocacy) will offer assessment options. However, not all are suitable for design work. In comparing options, we focus on approaches that are timely, actionable, and valid in the context of design. In no case are these `absolute' rules as there will be scenarios that require alternative approaches.

\textbf{Timely}---Assessments should be able to influence design decisions and be integrated into the design practice (ideally, they should be cheap and scalable). First, most visualization design activities are very time-constrained (i.e., deadline driven). Second, once deployed, the impacts of most affective visualization are likely short term (i.e., we would not expect a change in someone's value system 3 months after seeing a single visualization). Although some projects may run for extended periods and have many iterations of develop-deploy, most do not. Our preference will be for measurements that can be executed rapidly without significant involvement from the designer. For example, assessments that track longitudinal behavior and `grading' through a rubric or bio-physical measures~\cite{bamidis2017affective} are less desirable. Similarly, assessments that are unfamiliar are likely inappropriate (e.g., repertory grids~\cite{lemon1973attitudes} where participants create their own scales).

\textbf{Actionable}---As we have noted above, our goal is that assessments are useful for program-focused measures. We are much less interested in the impact of a visualization on individual viewers. That said, if we care about deploying different visualizations to different communities, we may consider groups of viewers. Ideally, the measure used should be sensitive enough so that we can isolate differences without having to recruit many participants. The differences (e.g., mean change in donations, difference in total donation, etc.) should allow the designer to make a decision of which visualization is better for their objectives. Measures that require huge populations will be less desirable. For example, randomized response methods use a coin flip to tell participants to either provide a fixed answer (e.g., `yes') or answer truthfully~\cite{horvitz1976randomized,warner1965randomized}\footnote{This provides plausible deniability to individual participants (around sensitive topics) and increases truthful answers. However, many participants are needed to remove the biased answers.}. 

\textbf{Valid}---Clearly, we would like assessments to satisfy all the constraints of good experiments: they should target the right population and should be reliable. They should have construct-validity (match the objective's construct), content-validity (sufficiently capture the objective), and predictive-validity (can we make an inference on the success of the objective from the measure?). None of these are unique to our domain. However, in the design case we leverage certain common features (e.g., access to multiple designs and potential for population randomization as a control).

\subsection{Assessments by Verb}
Having defined a few simple criteria and identifying the properties of our environment, we can now consider various evaluation approaches. We will work to map these to the specific objective structure (see Figure~\ref{fig:table}), roughly from the top to bottom row.

An important thing to note is that while assessment methods are often consistent along row (i.e., LO verb), the \textit{language} or \textit{scope} of the assessment may vary by the knowledge dimension (i.e., columns of Figure~\ref{fig:table}). These categories are Appraisal, Attitude, Value and Value System. An \textit{Appraisal} is a value judgment of a data fact (e.g., `the drought in Somalia is devastating a huge number of Somalis'). An \textit{Attitude} is a subjective opinion of an appraisal. For example, `the drought in Somalia is among the worst crises today.' \textit{Values} are more persistent beliefs that impact how the viewer interprets many situations. For example, \textit{equity} is a specific value through which we can view many situations such as humanitarian aid, hiring decisions, etc. Finally, \textit{Value systems} are groupings of values (e.g. a \textit{progressive} or \textit{conservative} value system). In describing the assessment strategies, we also discuss how they can be adapted to these categories.

\subsubsection{Assessments for \textit{Observe}}
\textit{Observe}, the most basic objective, aims for the viewer to `see' a space of possible beliefs (or appraisals, attitudes, etc.) exist. The target of this verb may be `free-standing' (e.g., `greenhouse gasses are bad for the environment'). As often, the target will be a spectrum of two things held in opposition (i.e., a debate). For example, one might hold the belief that `eliminating greenhouse gasses is the way to eliminate climate change' versus `eliminating aerosols is the way to eliminate climate change.' This space of views need not be a one-dimensional spectrum (e.g., there are many causes of climate change and which is most important can be debated). 

Observe objectives, whether targeted at appraisals, attitudes, values or value systems--are likely to be the simplest to measure using more standard assessment approaches (e.g.,~\cite{haladyna2004developing}). In this category, the designer wants the viewer to roughly model the belief space in question. For example, there is an argument over the factors that are leading to the crisis in Somalia (appraisal); or that individuals might believe in social responsibility through individual or government interventions (value). Assessments that can check recall, classification, or simple organization would likely work well here. For example, we could ask \textit{select all factors from the list that are impacting the humanitarian crisis in Somalia' with a list of true and distractor factors}. If the objective requires a deeper analysis, the question might be about properties or contrasts. For example, \textit{for each action in a list, indicate if they are the actions of someone who believes in individual social-responsibility or governmental social-responsibility.}

\subsubsection{Assessments for \textit{Position}}
\textit{Position} objectives go a step further beyond seeing the space of possible opinions one could take. These objectives specifically seek to place the viewer into a point in this space or that one point in this space is better than another (e.g., convincing the viewer to sit closer to the `greenhouse gas' side of the argument than the `aerosols'). 

Within the position level, designs can be assessed through measures of attitude and attitude change. Roughly, we would like the viewer to prefer one position over another. Many measures in this space are based on self-reports. Agreement or disagreement on whether one believes in a statement of appraisal, attitude, etc. may provide a signal of the viewer's placement. 

Attitudes can often be measured directly or indirectly through self-reports~\cite{smith2004instructional}. For example, a viewer can be asked directly if they agree that some position is favorable (e.g., \textit{Recycling is good}). They can also be asked indirectly by making judgments about the behaviors of others (e.g., \textit{In this scenario, Sally threw her can in the trash, do you agree with this behavior?}). 

Recall that a specific concern for visualization design is sufficient sensitivity, so we can see a result with fewer viewers. The formulation of survey statements can have a significant impact on the results. For example, agreement with the statements that we \textit{should allow speeches against democracy} versus \textit{should not forbid speeches against democracy} (ostensibly measuring the same attitude) results in different levels of agreement: 21\% vs 39\%~\cite{rugg1941experiments}. When comparing two visualizations, both statement forms may yield a reasonable ordering, but there is some risk if this is the only thing tested. Adding additional questions or survey instruments to test the same instrument may boost our confidence and mitigate the effects of a bad survey choice, while not adding too much burden on the participating viewer.

When self-reports are still untrustworthy, potentially for opinions that may have high social-undesirability, an alternative is to leverage the truth inducing approaches from psychology (e.g., error choice method~\cite{hammond1948measuring}) or economics~\cite{charness2021experimental}. Methods such as Choice Matching~\cite{cvitanic2019honesty}, Bayesian Truth Serum~\cite{prelec2004bayesian}, and Bayesian Markets~\cite{baillon2017bayesian}. Such methods often ask about the participants view's but also to guess or bet on what other people said. A correct guess may yield an additional reward. Such approaches are more incentive-compatible, but may require a certain sophistication and rationality in the respondents.

\subsubsection{Assessments for \textit{Strengthen}}
\textit{Strengthen} objectives build on \textit{position} even further by having the viewer more strongly `believe' or `disbelieve' in a particular viewpoint. Assessments that target the strength of a specific belief are often based on self-reports. There are various psychometric measures that have been extensively validated and can produce an indication of how strongly someone believes something. These include instruments such as Thurstone, Likert, Guttman, or Semantic Differential (see~\cite{coaley2014introduction,eagly1993psychology,krosnick2018measurement,mccoach13}). 

\begin{figure}[t]
     \centering
  \includegraphics[width=\linewidth, alt={}]{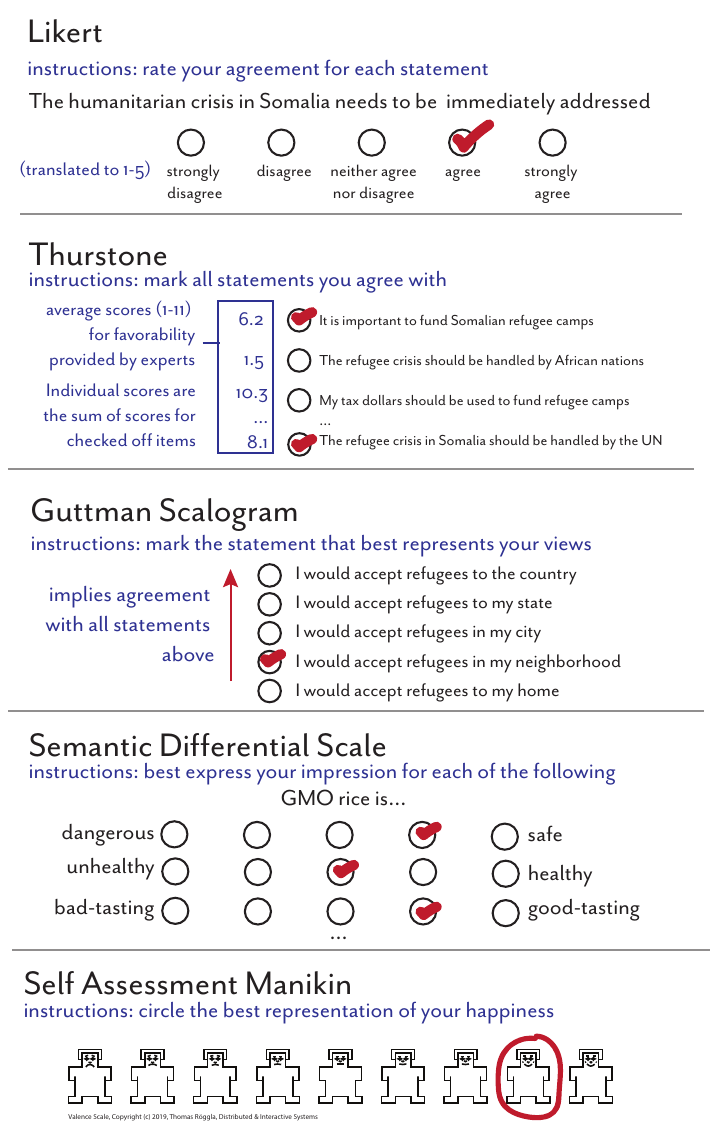}
  \caption{Examples of different scale types
  }
  \label{fig:scales}
  
\end{figure}

Figure~\ref{fig:scales} illustrates examples of scales. Some, like the Thurstone scale, require collecting favorability ratings by experts if constructed from scratch. However, there are many validated versions for specific topics that can be used directly or adapted (e.g., around social activism~\cite{turner2024development}). Some scales are more familiar (e.g., Likert) but may not be as easy to combine into one number. Scales such as the Guttman Scalogram are designed hierarchically so that agreement with one statement implies agreement with others. Semantic differential scales have the advantage of evaluating an opinion based on multiple criteria. In addition to written visual scales, visual scales such as the Self Assessment Manikin~\cite{bradley1994measuring} allow a participant to select a cartoon image representing their `feelings' on a subject.

In addition to the various scales above, there are a number of non-reactive and unobtrusive measures (see~\cite{webb2000unobtrusive}) that can be used to gauge opinion (both position and strength). For example, in an interactive system one could log engagement or time used as a proxy measure for an affective objective (e.g., see ~\cite{phan2019affective} for the case of a MOOC). Other proxy measures might be to provide the viewer with an opportunity to share, save, `like' or comment on a visualization. While most measures we have considered thus far are used in pre- and/or post-evaluation, instruments such as debate dials can be used when there is temporality to the presentation (e.g., the a video presentation of visualizations). Viewers can slide the dial in one direction or the other indicating (un)favorability while observing the material~\cite{saks2016dialed}.

\subsubsection{Assessments for \textit{Connect}}

\textit{Connect} objectives are those relating to situations where the viewer needs to contrast two or more beliefs that may not be in opposition to each other. For example, the designer might want to acknowledge that climate change and economic inequality are two pressing concerns for humanity, but that a point in the climate change space (e.g., `greenhouse gas-caused climate change' is more important than `globalization's impact on inequality'). These four objective types--observe, position, strengthen, and connect--are all `internal' objectives for the viewer. 

Connect objectives can be studied similarly to Strengthen and Position objectives. They simply involve additional instruments that consider additional orthogonal beliefs. For example, we might consider all the humanitarian crises and test the viewer's attitude towards them in relation to some target (e.g., a crisis in Somalia). As with Position objectives, Connect objectives might use ranking measures to determine how important certain issues are to an individual. 

Advocacy evaluation in particular often considers priorities as a key measure~\cite{reisman2007guide,weiss1995nothing}. In evaluating these efforts, participants are asked questions of the form: \textit{how seriously (do you / does your community) take each of the following issues?} or \textit{how likely would it be for you to support state funding towards each?} Such metrics may be adaptable to other affective objective scenarios. One specific constraint of advocacy, which may implicate visualizations used for advocacy, is that efforts of strategic communication may be targeted at very specific people (e.g., politicians who vote a certain way). In such cases, assessment may involve a narrower population of `viewers' (e.g., interviews with bellwethers or legislative staff~\cite{reisman2007guide}).

\subsubsection{Assessments for \textit{Behavior}}
A final objective category, \textit{Behave}, is the externalization of beliefs into behaviors (e.g., donating, voting, publicly advocating for, etc.). Ultimately, many affective objectives are about behavior. A behavior goal is that people should \textit{do} something based on a belief--vote, call their representative, donate, recycle, stop smoking, etc. In most cases, it is difficult to observe a behavior during an assessment of the design. That is, it is not always possible to provide an opportunity for the viewer of the visualization to demonstrate the behavior. In other cases, the desired outcome is that people stop doing some behavior (e.g., stopping to smoke). Observing or measuring this kind of behavior may present its own set of challenges. Due to the challenges of direct observation, many rely on behavioral proxies such as self-reports to measure \textit{intention to behave}.

Behavioral intentions \textit{can} predict behavior~\cite{ajzen1975intention} assuming certain conditions are met. These include: specificity (\textit{I intend to donate \$5 to X}); stability of the intention (how likely is it my intention to hold in the future?); and volitional control (how much control over the behavior does the person actually have). Clearly, not all behaviors align well in this way. The result is that we often see a weak correlation between self-reports and behavior~\cite{dang2020self}. Again, because we are often comparing two or more visualization designs, knowing which leads to a higher intention to behave is better than no measure. However, it may not indicate the same performance once the visualization is deployed.

As with other objectives, unobtrusive measures might be used as proxy measures for behavior~\cite{webb2000unobtrusive}. For example, accepting a recycling sticker may indicate an increased chance of recycling. As with other measures, unobtrusive measures require validation and therefore may not be suitable for all design tasks or objectives.

Another option for measuring behavior is to create opportunities within the assessment framework that allow for the demonstration of behavior. As we illustrate in the following case study, donation is one such measure. In addition to being closely aligned with the actual objective of interest (i.e., donation), this approach is incentive-compatible and preferable to self-reports. Not all behaviors can be simulated this way, although proxy behaviors (e.g., donation to a candidate) are correlated to other behaviors or beliefs (e.g., voting behavior or policy preferences~\cite{bonica2019donation}).

\subsection{Measuring Emotions and Other Theories}

An astute reader may notice that despite our focus on `affective'--which is often associated with \textit{emotion}--neither our taxonomy nor assessment approaches focus on emotion as a central concern.  In the context of \textit{research}, where mechanistic explanations are important, there are many instruments to assess emotional response~\cite{ekkekakis2013measurement,Meiselman2021mm}. Emotional responses fall into a broader class of constructs that relate to affective learning, often within a theoretical framing in psychology, persuasion theory, marketing, and so on (e.g.,~\cite{berlo,eagly1993psychology,elmo,falkheimer2022research,o2015persuasion,wells2014measuring,zimbardo1991psychology}). For example, the Elaboration Likelihood Model (ELM)~\cite{petty2012communication} looks at attitude change through the `route' by which that change can be induced (i.e., a central, deep reasoning route, or peripheral, more emotional or intuitive assessment). The model also provides an argument for what constitutes a strong (eliciting mostly positive reactions when thought about deeply) or weak message (mostly negative) in terms of elaboration. 

Designers could use knowledge of emotion's role in affective learning and apply that to their own design. 
For example, a designer may know that rage can be manipulated to induce donation~\cite{chapman2022rage} and may shy away from the colors that induce positive reactions and select those that induce disgust~\cite{kaya2004relationship}. However, the impact of a color choice in a \textit{specific design} on emotion and then donation may be impossible for a designer to practically determine.

In some ways, measuring emotion would simplify the designer's task. However, as we noted earlier, our model of affective objectives is that they should target things that can be learned, and thus retained. Emotions, although they influence behavior and learning, are often fleeting (e.g.,~\cite{GARRETT200139}). Beliefs, when they can be influenced, tend to be persistent (e.g.,~\cite{sears1999evidence}). A particular focus on emotion may not provide the timely, actionable, and valid assessments that would benefit design.

\section{Case Study}
\label{section:study}
In this case study, we showcase how to use the guidance in this paper to select and implement assessments based on affective learning objectives. We use the topic of a humanitarian crisis in Somalia, as it is a topic that most people do not have extensive prior knowledge about and has impact for affective learning objectives around attitudes, beliefs, and behaviors. Somalia experienced internal displacement and food insecurity as a result of a drought that affected 7 million people. We illustrate how learning objectives and assessments can be used in a real-world context by simulating our roles as designers working in collaboration with UN OCHA (the United Nations Office for the Coordination of Humanitarian Affairs). 

For this project, we are tasked with building an advocacy video~\footnote{While this is a simulated case study, we used reports and assets published by UN OCHA to create our materials. Some materials and data originally from a research project about data visualizations and statistical numbing.}. In this hypotethical design, UN OCHA has multiple documents about the crisis that are the `seed' to the videos. Some focus on individual stories~\cite{UNOCHA_2022} and other on data~\cite{UNOCHA_2023}. The target audience of this design project is members of the public who are unaware of the crisis.

Our design process follows a fairly conventional strategy: (1) identifying the goals for the project, articulated as learning objectives; (2) create several designs that could reasonably achieve our goals; and (3) choose an assessment for each goal by using the framework and guidelines in Section~\ref{section:assessments}. For this case study we demonstrate the implementation of the assessments through a crowdsourced user study on Prolific. Finally, we discuss the results of the survey and how we could use the information to inform our design process. 

\subsection{Affective learning objectives}
In our design process, we created these objectives. We infer these based on a close reading of materials from UN OCHA (e.g.,~\cite{UNOCHA_2022,UNOCHA_2023}).

\begin{itemize}[noitemsep,leftmargin=*]
    \item LO-A: The viewer will observe the humanitarian crisis in Somalia.
    \item LO-B: The viewer will believe that the crisis in Somalia is important.
    \item LO-C: The viewer will believe that they are able to personally make a difference in helping people in Somalia.
    \item LO-D: The viewer will believe that they have a moral responsibility to help people in Somalia.
    \item LO-E: The viewer will practice social responsibility and donate money to the Somalia Humanitarian Response Plan.
\end{itemize}

These objectives capture a range of LOs from our taxonomy (observe, strengthen, and behave).

\subsection{Visualization designs}
As designers, we created our artifact in the form of a short video that builds on existing UN OCHA materials. The existing materials give us many design assets (text, visualizations, maps, photographs, etc.). The fund's annual report (Humanitarian Response Plan~\cite{UNOCHA_2023}) significantly emphasizes data and visualizations. The report is published each year by UN OCHA and provides detailed information on the crisis and data on relief efforts (111 pages in the 2023 version we used). Critically, it contains many data visualizations on the current situation in Somalia, including maps, bar charts, donut charts, bubble/area charts, icon arrays, stacked bar charts, and line charts. 

An alternative document is a long-form humanitarian essay with stories about individual Somalis and Somali families~\cite{UNOCHA_2022}. The essay, titled ``Somalia: Hope fades as famine looms'', had numerous high-quality photographs, a significant amount of written content to use, and the content was compelling and emotional. The essay had a link to a donation site, suggesting that this would be effective content to achieve our affective goals. 

We identified three main topics of overlapping content in the Humanitarian Response Plan and the essay (acute food insecurity, internal displacement, and response efforts). These formed the core of the produced videos. In addition, we used written content from the foreword of the report, which summarized key points about the data. This became video voice-over when data visualizations were shown. 

The challenge for us in achieving the LOs is that we have alternative starting points \textit{and} conflicting information on what combination of visualizations, data, and narratives will work best.

\textbf{Data-Focused Content}---There is conventional wisdom in design~\cite{rall2016data} and research~\cite{markant2022can} that visualizations can shape belief. This can be achieved through a focus on data~\cite{Pandey_Manivannan_Nov_Satterthwaite_Bertini_2014} or in meta-visualizations (e.g., `fundraising thermometers'~\cite{fundtherm}). Given this, a natural video format would leverage the 2023 report's visualizations and text. In cases where we needed to supplement figures from the report (e.g., the text made a strong point but there was no visualization), we leveraged visualizations from news outlets (e.g., The Economist~\cite{EconomistSomalia}). The visualizations were across sources were well constructed and clearly communicated the need for humanitarian aid in Somalia. They made a logical argument about why a person should care about, and donate towards, the humanitarian crisis. 

\textbf{Human-Focused Content}---As designers we are aware that data visualizations can also be less effective in increasing prosocial behaviors than just highlighting the people in need. Several studies point out possible problems in focusing on data. For example, the identifiable victim effect is the effect where empathy is highest for an identifiable, determined, or known person~\cite{Kogut_Ritov_2005, small2003helping, Vastfjall_Slovic_Mayorga_Peters_2014}. 
Empathy is highest for a single individual, and every additional individual is valued at a lesser value than the one before. Even just the move from one person in need to two people in need results in compassion fade~\cite{Vastfjall_Slovic_Mayorga_Peters_2014}.  This trend of compassion fade continues in even larger numbers with psychic numbing, the phenomenon of people's decreasing empathy reactions to large numbers of suffering people~\cite{Slovic_2007, Lifton_1982}. 

Another drawback to showing the entire scope of the problem is that it can induce pseudo-inefficiency, where people think that their potential action will not be useful since it is just a drop in the bucket~\cite{vastfjall2015pseudoinefficacy}. 

Photojournalism portrays humans in their work~\cite{Lewis_2020, Zeeberg_2016}. Faces remind us that the data set is made up of real humans; designers should ``show what the data are about''~\cite{Rost2017}. This strategy is powerful because it is easier to form a mental image of and relate to a single person compared to an abstract dataset of people. An example of this is the photograph of Alan Kurdi, which sparked empathy for Syrian refugees that had not been elicited in previous news coverage~\cite{Demir_2017}.  

Considering the disadvantages of compassion fade and pseudo-inefficiency and advantages of photographs, the 2022 article would potentially be a better model for the video. Thus, our second video design focused on stories and photographs of impacted Somalis. We use both the visual content (photographs) and written content in this video. The written content became the narrative voice-over in the video while the photographs are being shown. 

\textbf{Balanced Content}---Because there may be benefits to both data-focused and human-focused approaches, a reasonable design hypothesis is that the combination is the `best of both worlds.' To test this, we built a third video that attempts to integrate content from both videos.

One design strategy is to redesign the data visualizations to incorporate elements of \textit{anthropographics}. 
Anthropographics are data visualizations that are designed to evoke feelings in the audience, such as empathy~\cite{Morais_Jansen_Andrade_Dragicevic_2020}.
There are various design strategies of how and what to visualize when visualizing data about people~\cite{Boy_Pandey_Emerson_2017, Morais_Jansen_Andrade_Dragicevic_2020}. Past work has identified seven dimensions of anthropomorphic design: granularity, specificity, coverage, authenticity, realism, physicality, and situatedness~\cite{Morais_Jansen_Andrade_Dragicevic_2020}. 
Various data visualization designs in this space have been investigated to see their effect on prosocial feelings and behaviors~\cite{Boy_Pandey_Emerson_2017, Campbell_Offenhuber_2019, Morais_Dandara_Sousa_Andrade_2020, Morais_Jansen_Andrade_Dragicevic_2021, Liem_Perin_Wood_2020}. However, these designs have been suggested to only have a negligible effect~\cite{Morais_Jansen_Andrade_Dragicevic_2021}. 

Given the prevailing evidence, we opted to more directly integrate the stories and photographs of the second video with the data of the first video.  
To create this combination video, 50\% of the content was data visualization and 50\% was individuals' stories.

In all, we constructed three video designs: data narrative, human narrative, and a mixed narrative combination of both. All three videos were matched for time and key topic coverage (the three main topics identified). More information about the videos can be found in the supplemental materials. With the hopes that the `best of both worlds' approach would speak to the widest range of viewers, we anticipated that it would result in the best `outcomes.'

\subsection{Selecting assessments}
In selecting the assessments for this project, we considered the pros and cons of methods that met our criteria (see Section~\ref{section:assessments}).

\begin{figure*}[t!]
    \centering
    \includegraphics[width=0.7\textwidth]{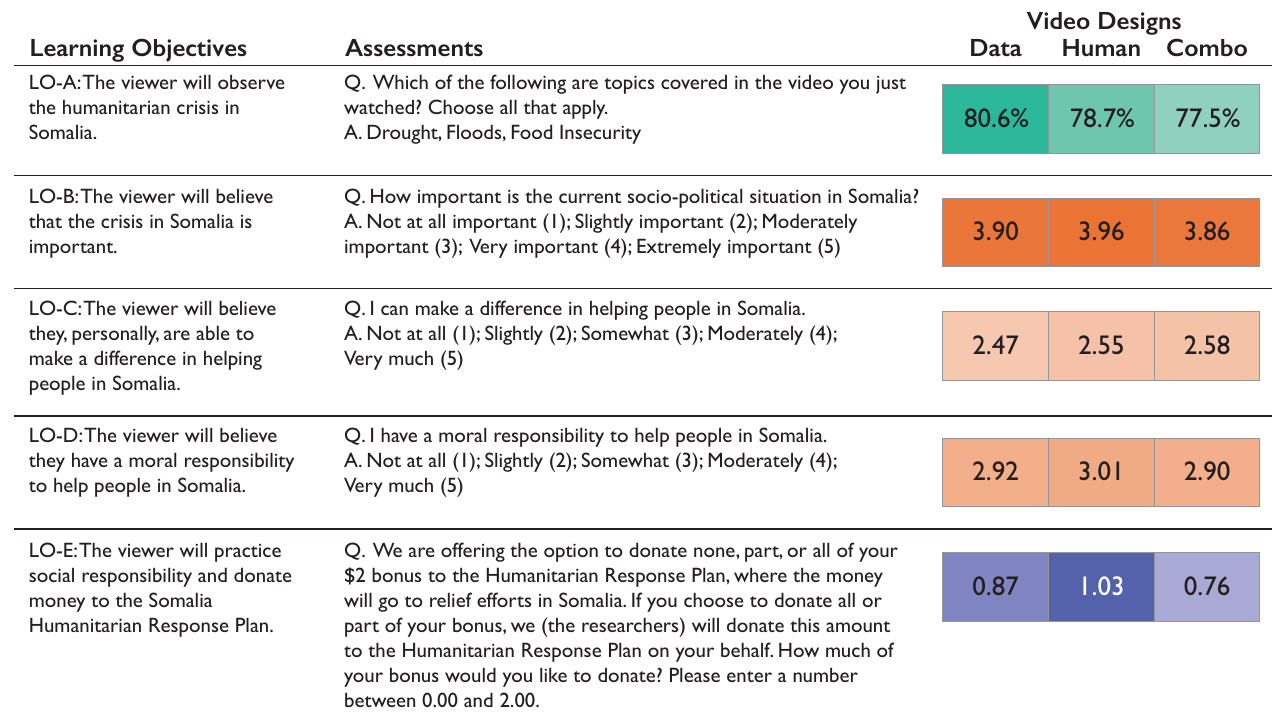}
    \caption{Our case study evaluated five affective learning objectives with five corresponding assessments for three video designs.}
    \label{fig:case-study}
    
\end{figure*}

\subsubsection{LO-A}
We start with selecting an assessment for LO-A: \textit{The viewer will observe the humanitarian crisis in Somalia}. This objective falls at the most basic level of the taxonomy (Observe/Appraisal). Based on our guidance, we opt for a standard cognitive assessment. In this case, we ask the viewer to recall details about the humanitarian crisis in Somalia. This is a basic question that will allow us to assess if the viewer has perceived the basis of the affective topic. 

\subsubsection{LO-B}
For LO-B (\textit{the viewer will believe that the crisis in Somalia is important}) the LO verb, \textit{believe}, falls under the category of \textit{strengthen} verbs\footnote{In this case, the Strength LO also measures Position because it is hierarchical. That is, if we assess the strength of the viewer's belief, we will also be able to tell what their Position is: do they agree that the crisis in Somalia is important?}. The `noun' part, \textit{the crisis in Somalia is important}, could be considered an \textit{attitude}. At this level of the taxonomy, the options for assessments are straightforward. We could ask a simple self-report question, e.g. ``\textit{How important is the current socio-political situation in Somalia?}'' A more complex truth-inducing approach is likely unnecessary\footnote{For example, we could also ask \textit{How important will} other \textit{people say the situation is?} and then apply a Bayesian Truth Serum analysis~\cite{prelec2004bayesian}}, as we have no reason to believe the topic is sensitive or that participants might lie. 

For our assessment, we can opt for a simple approach and simply ask: ``How important is the current socio-political situation in Somalia?'' 
For the scale, someone's attitude could fall on the spectrum from thinking it is not at all important to thinking it is extremely important. To get more granularity on where someone might fall in this range, we use a Likert scale as our multiple-choice options: Not at all important, slightly important, moderately important, very important, and extremely important. 

\subsubsection{LO-C and LO-D}
LO-C (\textit{the viewer will believe they personally are able to make a difference in helping people in Somalia}.) and LO-D (\textit{the viewer will believe they have a moral responsibility to help people in Somalia.}) both have \textit{strengthen} verbs and \textit{attitude} nouns. For assessment, we could use a self-report with scales or proxy engagement measures. Because each of our video designs is the same length, measuring a proxy engagement measure such as time spent viewing would not be an effective measure. Instead, we opt for the self-report question with a Likert scale.

One option for attitudes in particular is to modify an existing validated scale of attitudes. For example, Liem et al. (2020)~\cite{Liem_Perin_Wood_2020} used the questions from the European Social Survey~\cite{ESS} to measure values towards immigration. In our case, we can use a validated question from a related research study~\cite{Small_Loewenstein_Slovic_2007}. By using an assessment that has already been developed, we can have more confidence that it will be actionable, by being sensitive enough to show differences between different designs and has high validity. 

We asked participants to rate the following sentences.
\begin{itemize}[noitemsep,leftmargin=*]
    \item I have a moral responsibility to help people in Somalia. 
    \item I can make a difference in helping people in Somalia.
\end{itemize}

Because we are evaluating a Strengthen learning objective, we can use a Likert response scale for our multiple-choice options: Not at all; Slightly; Somewhat; Moderately; Very much.

\subsubsection{LO-E}

LO-E, \textit{the viewer will practice social responsibility and donate money to the Somalia Humanitarian Response Plan}, is the most challenging objective to assess. This learning objective falls under the category of behavior verbs and a value noun. The goal is to get the viewer to change their behavior related to a value, in this case, \textit{social responsibility}. Specifically, we want the viewer to take the action of donating money to the Somalia Humanitarian Response Plan.  

There are several ways to assess a behavior objective. Ideally, we might measure the direct behavior: the amount of money donated. However, if donation rates are low, we may need many participants. An alternative is to integrate donations directly into the protocol. For example, previous research gave participants 5 dollars as participation compensation and then gave them the option to donate some portion to a charity~\cite{Small_Loewenstein_Slovic_2007, small2003helping}. This approach is attractive as it is incentive compatible. However, this can be more difficult to implement by increasing the cost of data collection.

An alternative is to measure a proxy for the behavior, such as a self-report of behavior intentions. For example, we can create hypothetical donation allocation questions: either telling participants that \textit{'someone' was willing to make a \$10 donation in their name}~\cite{Boy_Pandey_Emerson_2017} or that they were to allocate a charity's hypothetical \$100 to different groups of migrants~\cite{Morais_Jansen_Andrade_Dragicevic_2021}. A hybrid between actual donation behaviors and donation intentions is to enter participants in a chance of winning an additional \$100 and asking them what amount of the bonus they would donate if they won~\cite{lindauer2020comparing}. However, past research has indicated a bias in self-reports of donations or likely donations~\cite{bekkers2011accuracy, lee2011dealing}, and concluded that there is a difference between hypothetical and real donations~\cite{maier2023revisiting}. 

For our case study, we decided to measure the direct behavior of \textit{actual} donations instead of a proxy behavior intention. One reason for this is that we wanted our assessment to have high sensitivity, and the direct behavior measurement is more likely to show differences between the video designs compared to if we used a self-report question.  In our case, we concluded that real donations were worth the cost. At the end of our survey, we give our participants the opportunity to donate none, part or all of a granted \$2 bonus to Somalia. Participants were able to donate an amount between \$0.00 and \$2.00. We note that we were limited in the \$2 bonus amount, as the Prolific system requires us to pay the participants for the survey to approve their work, but was not restrictive in whether or not we distributed the bonus. The bonus incentive is real money that the participant would receive if they chose not to donate. 

One experimental consideration is when to run assessments in relation to the video. A pre-- \textit{and} post-test has the advantage of identifying both absolute (A and B are better than baseline) and relative (A is better than B) `learning gains.' Focusing only on post-tests enables relative comparisons and is simpler to execute. In our case, we can largely use the post-test-only approach. However, to demonstrate absolute gains for LO-B, we employed a simplified pre-test survey.

\subsection{Running the Assessment}
As a demonstration of the design, we implemented our assessment via a survey on Prolific. We used the Prolific platform because it would be quick and scalable, giving us timely feedback about our designs that could be integrated into our design process. If we actually working with a partner such as UN OCHA, we could implement an A-B test on their website, showing real viewers different versions of our content. If we wanted to evaluate our designs before they are published, a crowdsourced study will be effective. Participants in our assessment earned a base compensation of \$2.80 with a bonus of \$2.00 for a total of \$4.80. On average, the survey took about 20 minutes to complete\footnote{In addition to the questions described above, we also measured several other constructs for a different research question. These are not analyzed in this paper.}.

\subsubsection{Results}
We analyzed data from 502 participants who participated in our survey, roughly split into three groups (173 data narrative, 164 human narrative, 165 in mixed narrative). For a summary of our assessment questions and results, see Figure~\ref{fig:case-study}. Naturally, running a study at this scale may not be strictly necessary for most design activities. Even `suggestive' evidence from a far smaller study (on the size of a pilot) may be sufficiently actionable for a designer. We briefly describe the results and their implications for the designer's choice. While the results are interesting, we emphasize that our focus is on using this as a case study to construct, execute and interpret affective assessments. They are specific to our three narrowly defined design alternatives.

First, we assessed whether participants observed the factors of the humanitarian crisis (LO-A). The data video design performed the best, with participants scoring at 80.6\% accuracy on this assessment. The human video design (78.7\%) and the combination video design (77.5\%) were not far behind. Overall, the scores for the first assessment were similar across all three video designs.

Judgments of importance were similar across all three designs (LO-B). On a scale of 1 (Not at all important) to 5 (Extremely important), the average importance ratings after seeing the video were 3.90 for the data video design, 3.86 for the combination video design and 3.96 for the people video design (moving from a pre-exposure baseline of 2.8). Thus, we see absolute learning gains but little relative differences. For LO-C and LO-D, participants were asked to rate agreement with the corresponding attitudinal statements (on a scale of 1 to 5). In general, the participants felt moral responsibility (M=2.9) slightly more than efficacy (M = 2.5), but none differed significantly by video design.

These first three objectives indicate that the LOs are achieved at roughly the same levels across all designs. However, for what is probably the most important objective (LO-E), the assessment shows there is a difference. The human video design elicited the most donations, with an average donation of 1.03 dollars. The combination video design garnered the least donations, with an average of 0.76 dollars. The data video performed somewhere in the middle (average of 0.87 dollars). 

As designers, what are we to conclude from this? For our first four learning objectives, all three designs resulted in similar results. Naturally, it is possible that our assessments were not sensitive enough to measure the differences between our conditions. If we were to re-evaluate our designs, we could reassess measures from Figure~\ref{fig:table} and opt for different assessments (e.g., a truth-inducing approach).

However, we see more variation between designs for the final assessment, donations. These results indicate that human-driven video design performs the best. Our `best of both worlds' video design (showing both data visualizations and personal narratives) actually resulted in the lowest average donation. The increase from \$0.76 to \$1.03 might not seem like much in the context of the survey. However, recall that the maximum donation amount was capped at \$2.00. A 36\% increase in donations could result in a meaningful amount when scaled up to an international organization like UN OCHA.

However, this result does not mean that the other videos are not useful. In this case study, we have exclusively targeted affective objectives. In reality, many communicative forms have multiple uses (e.g., to solicit donations from the public \textit{and} inform a board of directors). The design might call for both affective \textit{and} cognitive objectives (e.g., \textit{recall the number of deaths} or \textit{recall the number of clients served by the NGO}). In this case, an entirely photo-driven video would not work. The designer might opt for the data-driven video in that case. Although resulting in fewer donations (the human-only form), it would support the cognitive objectives.

Executing the assessment allows the designer to confirm (or refute) their hypothesis. It also provides useful data in the context of wicked design problems that enabled decision making. In addition, the designer can begin to develop some intuition on what works in this specific case (rather than relying on research studies for other contexts). In our case, we now have some evidence that data visualizations might numb an emotional response compared to stories and photos of individuals in need--suggesting the presence of compassion fade or statistical numbing in data visualization. While not definitive, the evidence might go against designer intuition and ultimately lead to better designs. The experience might also force thinking through alternative strategies beyond video (e.g., interactivity, scrollytelling, personalization, and other dynamic design choices to engage their audience and be more persuasive) or visualization approaches (e.g.,  anthropographic~\cite{Morais_Jansen_Andrade_Dragicevic_2020}).

\section{General Discussion} 

Past work has demonstrated that working through the creation of learning objectives and assessments will often yield better visualization design decisions~\cite{leerobbins22}. However, producing good objectives and assessments in the affective domain is challenging. The work presented here provides some guidance on how a designer can utilize this framework. 

One take on assessment---perhaps a controversial one---is having a good experimental design is desirable, but may not be critical for visualization \textit{designer} decision making. For example, it would be ideal that the sampled population gave us enough power and ensured the detection of statistically significant differences under any effect size. We also would rather not select a visualization design that is much worse than another because of some experimental design error. Some care is naturally warranted. However, the reality is that if the effect size is that small, it will make very little difference to the designer. At worst, an observed difference will be statistically meaningless and one design will be no worse than another. A corollary to this point is that not all affective visualizations need to perform complex assessments with external viewers. However, we believe that even reasoning through how a visualization \textit{might be assessed} would be helpful to a designer for selecting one design over another in a more principled manner. 

Despite our advocacy, it is worth acknowledging that even the best assessment possible---built to evaluate the most well articulated objective through the very best visualization design---may still not demonstrate a measurable change in belief or attitude. Belief instability, pliability, or centrality~\cite{crano2011attitudes,Rokeach68} may all impact whether a change can be made by a visualization, or any kind of communication. That said, there are many cases where long term exposure to \textit{many} forms of communication---including visualizations---can lead to real changes. A set of tools that can be applied to singular interventions or larger programs is valuable.

Finally, we emphasize that the focus in this paper is on defining objectives and assessment tools for designers when they are \textit{in the act of building} communicative visualization. However, this does not mean that the suggested assessments will not work in other scenarios (e.g., in critique or academic research). However, these scenarios may have different goals leading to additional assessment options. For example, the proposed methods may be effective in academic studies of affective visualizations. However, in this context, the research questions may focus more significantly on mechanisms and generalizations, so the objectives will focus on smaller or different constructs. The suggested assessments are also likely to work in larger design projects (e.g., creating a full advocacy website with some visualization elements), or in long term iterative design of visualizations after deployment. More `expensive' or time-consuming assessments (e.g., deeper interviews, instrumentation, etc.) may provide additional insights. It is also possible that the approaches we recommend may work in other communicative scenarios (beyond visualizations). However, as we observe in the case study, the specific features of visualizations--- for example, the rhetorical `competition' between pathos and logos and combinations of cognitive and affective objectives---may not be present in other formats (e.g., a simple advertising banner). The uniqueness of communicative visualizations inevitably requires a unique approach to design.

\section{Conclusion}

Learning objectives, both cognitive and affective, can help data visualization designers specify their goals: What do they expect from their viewers after engaging with a visualization? Unlike cognitive objectives, affective objectives are harder to specify and assess. That is, determining whether the viewer learned a specific data fact from the visualization is easier to determine than whether they have changed their belief on some topic. In this work, we contribute guidelines for considerations when choosing assessments when designing visualizations. We apply these guidelines in evaluating assessments from many domains (economics, advocacy, marketing, psychology). We conclude with a case study that demonstrates the creation of affective objectives (at different levels), the construction of assessments, and examples of their implementation. Assessing affective goals, while difficult, is not an impossible task. Although we do not expect designers to create perfect scientific experiments, we hope that this work enables them to create `good enough' assessments. Rather than relying on intuition on what works, which can often be wrong when affective objectives are concerned, this structure provides a mechanism to build designs that are objectively more effective.

\acknowledgments{%
    Thanks to Alison Siegler, Francis Gagnon, Gabrielle Merite, Dino Citraro, and Peiying Loh for their permission to reuse their images. The authors thank the NSF for their support of this work through NSF IIS-1815760. This work was supported in part by funding from the Rackham Graduate School at the University of Michigan.%
}

\balance
\bibliographystyle{abbrv-doi-hyperref}
\bibliography{88_bib}

\end{document}